\documentclass[twocolumn,preprintnumbers,amsmath,amssymb,superscriptaddress]{revtex4-2}
\usepackage{graphicx}
\usepackage{dcolumn}
\usepackage{bm}
\usepackage{color}
\usepackage{ulem}

\begin{document}
\title{Binding two and three $\alpha$ particles in cold neutron matter}
\author{H. Moriya}
\email{moriya@nucl.sci.hokudai.ac.jp}
\affiliation{Department of Physics, Hokkaido University, Sapporo 060-0810,
  Japan}
\author{H. Tajima}
\email{hiroyuki.tajima@phys.s.u-tokyo.ac.jp}
\affiliation{Department of Physics, Graduate School of Science, The University of Tokyo, Tokyo 113-0033, Japan}
\affiliation{Department of Mathematics and Physics, Kochi University, Kochi 780-8520, Japan}
\author{W. Horiuchi}
\email{whoriuchi@nucl.sci.hokudai.ac.jp}
\affiliation{Department of Physics, Hokkaido University, Sapporo 060-0810,
  Japan}
\author{K. Iida}
\email{iida@kochi-u.ac.jp}
\affiliation{Department of Mathematics and Physics, Kochi University, Kochi 780-8520, Japan}
\author{E. Nakano}
\email{e.nakano@kochi-u.ac.jp}
\affiliation{Department of Mathematics and Physics, Kochi University, Kochi 780-8520, Japan}
\date{\today}
\begin{abstract}
  We elucidate the fate of neighboring two and three $\alpha$ particles
  in cold neutron matter by focusing on an analogy between
  such $\alpha$ systems and Fermi polarons realized in ultracold atoms.
  We describe in-medium excitation properties of an $\alpha$ particle
  and neutron-mediated two- and three-$\alpha$ interactions
  using theoretical approaches developed for studies of cold atomic systems.
  We numerically solve the few-body Schr\"{o}dinger equation of $\alpha$
  particles within standard $\alpha$ cluster models combined with
  in-medium properties of $\alpha$ particles. We point out that
  the resultant two-$\alpha$ ground state and three-$\alpha$ first excited
  state, which correspond to $^8$Be and the Hoyle state, respectively,
  known as main components in the triple-$\alpha$ reaction,
  can become bound states in such a many-neutron background although
  these states are unstable in vacuum. Our results suggest a significance
  of these in-medium cluster states not only in astrophysical environments
  such as core-collapsed supernova explosions and neutron star
  mergers but also in neutron-rich nuclei.
\end{abstract}
\maketitle

\section{Introduction}

An $\alpha$ particle ($^{4}$He nucleus) has a significantly large binding
energy compared to other light elements
and hence can be an important ingredient in understanding
the structure of nuclei as well as the origin of elements. 
In light $N=Z$ nuclei,
the threshold energy for $\alpha$ particle disintegration 
becomes low and even comparable to the one-$\alpha$ separation energy,
which helps the $\alpha$ cluster structure to emerge
in the spectrum of such light nuclei
as predicted by the Ikeda diagram~\cite{Ikeda68}.

One of the most famous examples of the $\alpha$ cluster
structure is the first excited $J^\pi=0^+$ state of $^{12}$C,
which was originally predicted by Fred Hoyle~\cite{Hoyle54},
This state, which is often called the Hoyle state, is 
recognized as having a well-developed three-$\alpha$ cluster structure.
The existence of such cluster states plays a role in enhancing
the reaction rate at extremely low energies near the Gamow window.
The $^{12}$C element forms dominantly through a sequential reaction
in which a resonant two-$\alpha$ system, the ground state of $^{8}$Be,
absorbs another $\alpha$ particle 
via radiative capture process~\cite{Salpeter52}.
The accurate description of such $\alpha$ induced reactions
can impact astrophysically important explosive phenomena~\cite{Oertel17},
such as core collapse supernovae and
neutron star mergers, which have recently started to be
measured through gravitational waves~\cite{LIGO17}.

The importance of $\alpha$ clusters has 
extended from light nuclei to many-nucleon systems such as
medium-heavy nuclei and nuclear matter.
The role of $\alpha$ particles in supernova explosions
has attracted attention~\cite{Hempel11}.
Very recently, an interesting indication that
$\alpha$ clusters emerge in a surface region of 
medium-heavy mass nuclei have been obtained by a systematic 
measurement via the $\alpha$ knockout reactions~\cite{Tanaka21}.
These encourage us to study the formation and structure
of an $\alpha$ particle in dilute neutron-rich matter.
In Ref.~\cite{Nakano20}, three of the present authors 
(E.N., K.I., and W.H.) discussed the static properties of
an $\alpha$ particle in cold dilute neutron matter.
The effective mass of the in-medium $\alpha$ particle
is enhanced by the interaction with the neutron matter,
implying the possibility of binding the ground state resonance
of $^{8}$Be and the Hoyle state in such an extreme environment.
If realized,
these molecular-like `bound' states will take part
in the astrophysical reactions,
in addition to compact multi-$\alpha$ cluster systems,
  e.g., the ground states of $^{12}$C and $^{16}$O,
and thus should be incorporated
explicitly as ingredients of simulations of 
astrophysical nuclear processes~\cite{Roepke20},
which may affect the local abundance of 
the chemical elements.

In general, it is challenging to see how impurity particles 
behave in many-body backgrounds like a Fermi sea due to infinitely 
large degrees of freedom.  Nevertheless, this problem has been 
tackled in ultracold atoms theoretically and experimentally 
in terms of Fermi polarons~\cite{Chevy10,Massignan14,Schmidt18};
an impurity atom is dressed by excitations of majority Fermi atoms 
via interspecies interactions.  Quasiparticle properties of a single polaron
such as the effective mass have precisely been measured in 
experiments~\cite{Nascimbene09,Schirotzek09,Sommer11,Kohstall12,Cetina16,Scazza17,Yan19,Ness20,Fritsche21} 
and successfully described by various theoretical frameworks such as a variational method~\cite{Chevy06} 
and a $T$-matrix approximation~\cite{Combescot07}.
Moreover, fermion-mediated interactions between polarons
have also been observed experimentally~\cite{DeSalvo19,Edri20}.

In this work, we investigate the structure of two- and
three-$\alpha$ systems in dilute neutron matter of density lower than
about 1/100 of the saturation density $\sim 0.01\rho_0$ at zero temperature
and discuss their medium-induced stabilization by regarding each
$\alpha$ particle as a mobile impurity immersed in the neutron medium.
Analogy with Fermi polarons realized in ultracold atoms allows us to
utilize the results obtained for quasiparticle properties of 
a single $\alpha$ particle in neutron matter~\cite{Nakano20} by using 
Chevy's variational ansatz known to give
a quantitative description of Fermi atomic polarons.
To discuss the stability of two- and three-$\alpha$ particles
immersed in neutron matter,
moreover, we derive medium-induced two- and three-body interactions 
among polarons using a diagrammatic approach.
Once the effective Hamiltonian is set, the structure
of the in-medium two- and three-$\alpha$ systems 
can be accurately obtained from the solution of
the corresponding few-body Schr\"odinger equation.
This study offers the first quantitative 
evaluation of the energy and the pair density distribution
of two and three-$\alpha$ systems in cold neutron matter.

This paper is organized as follows.
The next section describes models of the in-medium multi-$\alpha$ systems.
Section~\ref{inducedint.sec} is devoted to the derivation
of induced two- and three-$\alpha$ interactions in a neutron Fermi sea.
Section~\ref{hamil.sec} gives the effective Hamiltonian for
the multi-$\alpha$ systems in the cold neutron matter.
The two cluster models employed in this study are
briefly described in Sec.~\ref{clustermodel.sec}.
Given the effective Hamiltonian, in Sec.~\ref{CG.sec},
we address how to solve the few-$\alpha$ Schr\"odinger 
equation precisely using the correlated Gaussian expansion.
Section~\ref{results.sec} presents our results.
The possibility of the medium-induced stabilization of the 
two- and three-$\alpha$ systems is discussed.
The conclusion and future prospects are given in Sec.~\ref{conclusion.sec}.

\section{Models of in-medium two- and three-$\alpha$ systems}

Let us proceed to construct models for the systems of
two and three $\alpha$ particles of bare mass $M$ in a dilute gas of 
neutrons of bare mass $m$ at zero temperature.
Since we are interested in $\alpha$ particles in astrophysical
environments where the temperature is higher than the neutron superfluid 
critical temperature~\cite{Oertel17}, we can safely assume that the 
neutron gas is in a normal state.
We are interested in astrophysical situations where $\alpha$
particles occur thermally rather than by external factors;
hence, the crust of very cold neutron stars is out of our scope.
We also ignore
the neutron-neutron interaction for simplicity.
Although we employ the zero-temperature results for the induced 
interactions among $\alpha$ particles and
the $\alpha$ effective mass as will be discussed below, such zero-temperature 
treatment can be justified when the temperature is below both the
neutron Fermi temperature $T_F=\frac{\hbar^2k_F^2}{2mk_B}$ and the cutoff 
energy scale $\frac{\hbar^2}{m_{r}r_0^2}\simeq 25$~MeV of the 
neutron-$\alpha$ interaction with the effective range 
$r_0=1.43$~fm~\cite{Nakano20}.
We finally remark that at sufficiently
high neutron densities corresponding to $k_F\gtrsim0.3$ fm$^{-1}$, 
a $p$-wave resonance ($^5$He) could be stabilized by the Pauli blocking
effect and emerge as a nuclear ingredient \cite{Roepke20,Nakano20}.
This possibility is another issue to be tackled with, but is beyond 
the scope of this work.

\subsection{Derivation of induced two- and three-body interactions
in cold neutron matter}
\label{inducedint.sec}

\begin{figure}[t]
\begin{center}
\includegraphics[width=\linewidth]{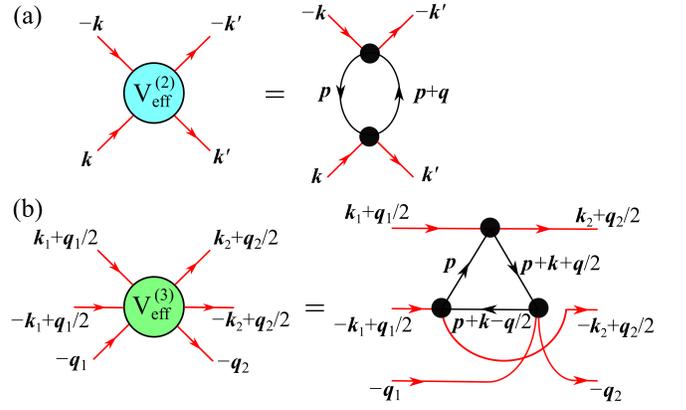}
\end{center}
\caption{
  Feynman diagrams in the center-of-mass frame of $\alpha$ particles
    that represent the induced (a) two-body interaction 
$V_{\rm eff}^{(2)}(\bm{q},i\nu_\ell)$ and (b) three-body interaction 
$V_{\rm eff}^{(3)}(\bm{k},\bm{q},i\nu_{\ell},i\nu_{u})$ among
$\alpha$ particles immersed in neutron matter.
For $V_{\rm eff}^{(2)}(\bm{q},i\nu_\ell)$,
the incoming (outgoing) momenta of $\alpha$ particles are given by 
$\bm{k}$ and $-\bm{k}$ ($\bm{k}'$ and $-\bm{k}'$).
For $V_{\rm eff}^{(3)}(\bm{k},\bm{q},i\nu_{\ell},i\nu_{u})$,
the incoming (outgoing) momenta of $\alpha$ particles are given by
$\bm{k}_1+\bm{q}_1/2$, $-\bm{k}_1+\bm{q}_1/2$, and $-\bm{q}_1$ 
($\bm{k}_2+\bm{q}_2/2$, $-\bm{k}_2+\bm{q}_2/2$, and $-\bm{q}_2$).
The internal solid lines denote the thermal Green's function of a neutron.
}
\label{fig:diagram}
\end{figure}

We start with diagrammatic derivation of the medium-induced two- and 
three-body interactions among $\alpha$ particles in a neutron Fermi sea. 
As depicted diagrammatically in Fig.~\ref{fig:diagram}(a),
the induced two-body interaction between two $\alpha$ particles
can be obtained up to leading order in $a$
as~\cite{Sheehy}
\begin{align}
&V_{\rm eff}^{(2)}(\bm{q},i\nu_\ell)
= -\left(\frac{2\pi \hbar^2 a}{m_r}\right)^2\notag\\
&\times \frac{k_BT}{\hbar^2}\sum_{\sigma=\uparrow,\downarrow}\sum_{\bm{p},\omega_n}G_{\sigma}(\bm{p}+\bm{q},i\omega_n+i\nu_\ell)G_{\sigma}(\bm{p},i\omega_n),
\end{align}
where $k_B$ is the Boltzmann constant,
$(\bm{q},i\nu_\ell)=(\bm{k}-\bm{k}',i\nu_{s}-i\nu_{s'})$
is the transferred four-momentum,
$\nu_\ell=2\ell \pi k_BT/\hbar$ is the bosonic Matsubara frequency~\cite{FW},
$G_{\sigma}(\bm{p},i\omega_n)=\left(i\omega_n-\xi_{\bm{p}}/\hbar\right)^{-1}$
is the thermal Green's function of a neutron
with energy $\xi_{\bm{p}}=\frac{p^2}{2m}-\varepsilon_F$ 
relative to the neutron Fermi energy $\varepsilon_F$, 
and $m_r=(m^{-1}+M^{-1})^{-1}$ is the reduced mass.
$a=2.64$~fm is the $s$-wave neutron-$\alpha$ scattering length~\cite{Nakano20}.
Taking the summation of the fermionic Matsubara frequency
$\omega_n=(2n+1)\pi k_BT/\hbar$~\cite{FW}, 
we obtain the induced two-body interaction as
\begin{eqnarray}
\label{eq2}
V_{\rm eff}^{(2)}(\bm{q},i\nu_\ell)=2\left(\frac{2\pi \hbar^2 a}{m_r}\right)^2\sum_{\bm{p}}\frac{f(\xi_{\bm{p}})-f(\xi_{\bm{p+q}})}{i\hbar\nu_\ell+\xi_{\bm{p}}-\xi_{\bm{p+q}}}.
\end{eqnarray}
In the low-energy limit $\nu_\ell=0$ at $T=0$,  Eq.\ (\ref{eq2}) reduces to
\begin{align}
\label{eq3}
V_{\rm eff}^{(2)}(\bm{q},0)&=
-\frac{mk_F}{2\pi^2\hbar^2}\left(\frac{2\pi\hbar^2 a}{m_r}\right)^2\notag\\
&\times\left[1+\frac{k_F}{q}\left(1-\frac{q^2}{4k_F^2}\right)\ln\left|\frac{q+2k_F}{q-2k_F}\right|\right].
\end{align}
Note that in the long wavelength limit ($\bm{q}\rightarrow 0$), Eq.~(\ref{eq3}) 
can be expressed by the compressibility 
$\kappa=\frac{1}{\rho^2}\left(\frac{\partial \rho}{\partial\mu}\right)$ 
of neutron matter as $V_{\rm eff}^{(2)}(\bm{q}\rightarrow\bm{0},0)=-\left(\frac{2\pi\hbar^2 a}{m_r}\right)^2\rho^2\kappa$.
By taking the inverse Fourier transformation of Eq.~(\ref{eq3}),
we obtain the well-known Ruderman-Kittel-Kasuya-Yosida (RKKY) form of the induced two-body 
interaction in the coordinate space as~\cite{VanVleck1962,Spielman2014,Nishida2009,Liao2020,Suchet2017}
\begin{align}
  V_{\rm eff}^{(2)}(\bm{r}_1,\bm{r}_2)
  &=\frac{m}{8\pi^3\hbar^2}\left(\frac{2\pi\hbar^2a}{m_r}\right)^2\notag\\
  &\times\frac{(2k_Fr)\cos(2k_Fr)-\sin(2k_Fr)}{r^4},
\label{RKKY.eq}
\end{align}
where $r=|\bm{r}_1-\bm{r}_2|$.

Moreover, as diagrammatically drawn in Fig.~\ref{fig:diagram}(b), the induced 
three-body interaction up to leading order in $a$ is given by~\cite{Tajima21}
\begin{align}
  &V_{\rm eff}^{(3)}(\bm{k},\bm{q},i\nu_{\ell},i\nu_{u})
  =2\left(\frac{2\pi\hbar^2 a}{m_r}\right)^3\notag\\
  &\times \frac{k_BT}{\hbar^3}\sum_{\sigma=\uparrow,\downarrow}\sum_{\bm{p},\omega_n}G_{\sigma}(\bm{p},i\omega_n)
G_{\sigma}(\bm{p}+\bm{k}+\bm{q}/2,i\omega_n+i\nu_{\ell})\notag\\
&\times G_{\sigma}(\bm{p}+\bm{k}-\bm{q}/2,i\omega_n+i\nu_{\ell}-i\nu_{u}),
\end{align}
where $\bm{k}=\bm{k}_1-\bm{k}_2$, $\bm{q}=\bm{q}_1-\bm{q}_2$ 
$i\nu_{\ell}=i\nu_{s_1}-i\nu_{s_2}$, and $i\nu_{u}=i\nu_{j_1}-i\nu_{j_2}$
are the transferred four-momenta.
In the low-energy limit ($i\nu_{\ell}=i\nu_{u}=0$),
the induced three-body interaction in the coordinate space can be obtained as
\begin{align}
  V_{\rm eff}^{(3)}(\bm{r}_1,\bm{r}_2,\bm{r}_3)&=\sum_{\bm{k},\bm{q}}V_{\rm eff}^{(3)}(\bm{k},\bm{q},0,0)
  e^{-i\bm{k}\cdot\bm{x}_1+i\bm{q}\cdot \bm{x}_2},
\end{align}
where $\bm{x}_1=\bm{r}_1-\bm{r}_2$ and $\bm{x}_2=\bm{r}_3-(\bm{r}_1+\bm{r}_2)/2$.
For simplicity, we employ
the contact-type three-body interaction whose coupling constant is given by
\begin{eqnarray}
  V_{\rm eff}^{(3)}(\bm{0},\bm{0},0,0)&=&2\left(\frac{2\pi\hbar^2 a}{m_r}\right)^3
  \frac{k_BT}{\hbar^3}\sum_{\sigma}\sum_{\bm{p},i\omega_n}[G_{\sigma}(\bm{p},i\omega_n)]^3\cr
&=&\frac{m^2}{\pi^2\hbar^4k_F}\left(\frac{2\pi\hbar^2 a}{m_{r}}\right)^3.
\end{eqnarray}
Thus, we obtain
\begin{align}
  V_{\rm eff}^{(3)}(\bm{r}_1,\bm{r}_2,\bm{r}_3)&=\frac{m^2}{\pi^2\hbar^4k_F}\left(\frac{2\pi\hbar^2 a}{m_{r}}\right)^3\delta(\bm{x}_1)\delta(\bm{x}_2).
\label{v3eff.eq}
\end{align}
 Note that to adopt the contact interaction (\ref{v3eff.eq}) is equivalent to the local density approximation.

\subsection{Hamiltonian for two- and three-$\alpha$ particles in cold neutron
matter}
\label{hamil.sec}

A single $\alpha$ particle immersed in cold neutron matter
has its mass $M$ changed into the effective mass $M^*$ 
by the interaction with neutrons in the medium.
This particle, dressed with neutron excitations, can be 
regarded as a polaron.
As a natural extension of the previous study on this polaron~\cite{Nakano20},
we consider two- and three-$\alpha$ systems immersed
in cold neutron matter.
As we shall see, the mass enhancement through $M^*$ acts to 
increase binding of these systems.
The explicit form of the Hamiltonian of the three-$\alpha$ system in
cold neutron matter is
\begin{align}
  H&=\sum_{i=1}^3 \frac{\bm{p}_i^2}{2M^*}-T_{\rm cm}\notag\\
  &+\sum_{i<j=1}^3 \left[U^{(2)}_{ij}
    +V^{(2)}_{{\rm eff}; ij}\right]+U^{(3)}+V^{(3)}_{\rm eff},
\end{align}
where the center-of-mass kinetic energy term $T_{\rm cm}$
is subtracted, $U^{(x)}$ $(x=2,3)$
denotes the $x\alpha$ potential in vacuum including the Coulomb term,
and $V^{(x)}_{\rm eff}$ is the induced $x\alpha$ interaction in the neutron
medium with the Fermi momentum $k_F$.
Note that $M^*$ and $V^{(x)}_{{\rm eff}}$ depend on $k_F$.
The $k_F$ dependence of $M^*/M$ is taken from Ref.~\cite{Nakano20}.
We take the neutron mass as $\hbar^2/m$= 41.47 MeV\,fm$^2$
and $M=4m$  to keep the consistency of the parameters
given in Ref.~\cite{Nakano20}.

Here we incorporate
$V^{(2)}_{\rm eff}$ and $V^{(3)}_{\rm eff}$
derived in the previous subsection into the Hamiltonian.
The original RKKY potential (\ref{RKKY.eq}) behaves as
$\sim r^{-1}$ at small $r$ and hence has a singularity at the origin. 
This is regularized by folding the harmonic oscillator type form
factor of the $\alpha$ particle associated with the nuclear force, 
$\left(\frac{8\nu}{3\pi}\right)^{\frac{3}{2}}e^{-\frac{8}{3}\nu u^2}$, 
which leads to
\begin{align}
  V^{(2)}_{\rm eff}(r)
  =V_{\rm RKKY}(r){\rm erf}\left(\frac{4}{3}\sqrt{\nu} r\right),
  \label{RKKYreg.eq}
\end{align}
where $\nu$ is also taken as 0.2675 fm$^{-2}$ in a way that is 
consistent with the width parameter of the $\alpha$ particle~\cite{Fukatsu92}.
Note that this range is shorter than the $\alpha$-$n$ scattering length $a$
and $1/k_F$ considered in this work.
It is reasonable to take the range of the induced three-body force
as the same as the one for the induced two-$\alpha$ interaction, which 
leads to
\begin{align}
  V_{\rm eff}^{(3)}(R)&=\frac{m^2}{\pi^2\hbar^4k_F}\left(\frac{2\pi\hbar^2 a}{m_{r}}\right)^3N_\nu e^{-\frac{16}{9}\nu R^2}
\end{align}
with the normalization constant of the form factor 
$N_\nu=\left(\frac{16\nu}{3\pi}\right)^{3}$.
Note that $R^2=(\bm{r}_1-\bm{r}_2)^2+(\bm{r}_2-\bm{r}_3)^2+(\bm{r}_3-\bm{r}_1)^2=\frac{3}{2}x_1^2+2x_2^2$, which is symmetric in any particle exchange.
Since $a$ is positive, the induced three-$\alpha$ potential is always 
repulsive; its strength is inversely proportional to $k_F$.

\begin{figure}[t]
\begin{center}
\includegraphics[width=\linewidth]{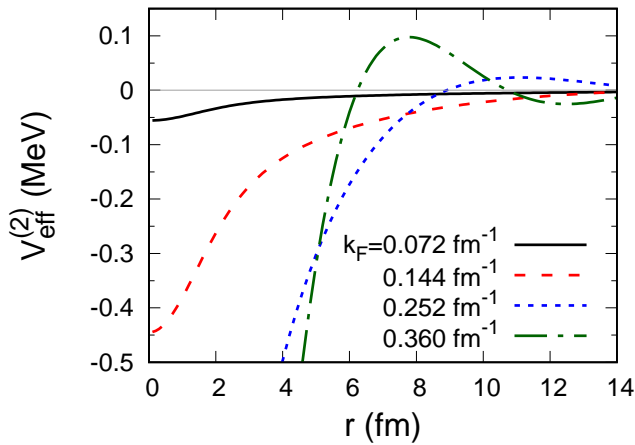}
\end{center}
\caption{Induced two-body interaction $V_{\rm eff}^{(2)}(r)$ 
  as a function of the $\alpha$-$\alpha$ distance $r$
  for various values of the Fermi momenta $k_F$ of cold
  neutron matter. The thin horizontal line indicates zero.}
\label{fig:induced2bf}
\end{figure}

Figure~\ref{fig:induced2bf} plots the $k_F$ dependence
of the induced two-$\alpha$ potential, Eq.~(\ref{RKKYreg.eq}).
At short distances, this two-$\alpha$ interaction is attractive,
leading to more stability of multi-$\alpha$ systems.
The range  of such attraction increases with $k_F$,  while,
for sufficiently large $k_F$, some oscillatory behavior appears,
a feature reflecting the Friedel oscillation associated with the 
presence of the neutron Fermi surface.
The induced three-$\alpha$ interaction,  on the other hand, is 
repulsive and weakens as $k_F$ increases. 
The optimal stability of the three-$\alpha$ system
can thus be realized at a certain $k_F$ that is determined
in balance with the purely repulsive induced three-$\alpha$ potential.

The difference of the effective mass from the bare mass,
together with the induced interactions, can crucially affect 
the relative motion between $\alpha$ particles.
In this work, we consider each $\alpha$ particle to be
a structureless particle but treats the Pauli principle
in the interaction between $\alpha$ particles in two different ways.
Both potential models well reproduce the empirical $\alpha$-$\alpha$ 
scattering phase shift.  Although it is difficult to obtain empirical
information on the closest motion, they are known to give different 
results for the internal region of the relative wave function.
See, e.g.,~\cite{Pinilla11, Arai18,Moriya21}
for some examples in light cluster systems.
We utilize such two potential models to evaluate
the uncertainty that comes from  model choice.

\subsection{Multi-$\alpha$ cluster models}
\label{clustermodel.sec}

We start with a standard type of $\alpha$ cluster model that 
assumes a shallow and repulsive potential.
As $U^{(2)}$, we employ the Ali-Bodmer (AB) potential~\cite{AB}
(Set a$^\prime$~\cite{Fedorov96}), which
reproduces the $\alpha$-$\alpha$ scattering phase shift
and produces the $s$-wave $^{8}$Be ($0_1^+$) 
resonance position with 0.093 MeV,
a value close to the empirical one 0.092 MeV~\cite{Tilley04}.
Note that we get 0.086 MeV in the present calculation
because of the use of the different mass parameter 
of an $\alpha$ particle ($M=4m$).
The AB potential is $l$-dependent
and its explicit form is
\begin{align}
  U_{\rm AB}(r)& = \left(125\hat{P}_{l=0}+20\hat{P}_{l=2}\right)
  \exp\left(-\frac{r^2}{1.53^2}\right)\notag\\
&-30.18\exp\left(-\frac{r^2}{2.85^2}\right), 
\label{2apot1.eq}
\end{align}
where the energy and length are given in units of MeV and fm,
and $\hat{P}_{l}$ is the projection operator onto
the relative angular momentum $l$.
This potential is so shallow that no bound state appears. 
The Pauli principle in the interaction
between $\alpha$ particles is simulated
by the first repulsive term of the potential.
It is known that the empirical energies of states close to the
threshold energy of the three-$\alpha$ system
are not well reproduced by the two-body interaction alone~\cite{Suzuki02}. 
 Then, one often introduces a phenomenological
 three-$\alpha$ potential as $U^{(3)}$,
  which only has a single Gaussian attractive term~\cite{Ishikawa13}.
Because of such simplicity,
a similar sort of potential model
has often been used to describe astrophysically
important reactions~\cite{Ogata09, Nguyen12, Ishikawa13, Akahori15, Suno16, LHP20,LHP21}.
This three-$\alpha$ interaction, together with the two-$\alpha$ one,
leads to the Hoyle state energy of 0.38 MeV with respect to the
three-$\alpha$ threshold,
which perfectly agrees with the empirical Hoyle state energy
~\cite{Ajzenberg90}.

Another standard cluster model employs a deep attractive potential,
which accommodates three redundant bound states
$\phi_{n_fl_fm_f}$ with $(n_f,l_f)=(0,0), (1,0),(0,2)$
that are forbidden by
the Pauli principle in the interaction between two $\alpha$ particles.
When the two- and three-$\alpha$ equations are solved,
the orthogonality condition
to be imposed for an $N$-$\alpha$ system reads
\begin{align}
  \sum_{i<j=1}^N\sum_{nlm\in f}\left|\left<\phi_{nlm}(ij)|\Psi\right>\right|^2=0,
\end{align}
where $f$ and $\Psi$ denote the Pauli forbidden two-$\alpha$ bound states and 
the eigenstate of the system, respectively.  That is why this kind of model 
is called the orthogonality condition model (OCM)~\cite{OCM1,OCM2,OCM3},
 which has often been used as an alternative to the microscopic
cluster model and been successful in describing
the $\alpha$ condensed states predicted for
$^{12}$C and $^{16}$O~\cite{Yamada05, Funaki08}.
As demonstrated in Ref.~\cite{Schmid61},
the low energy $\alpha$-$\alpha$ scattering phase shifts are
well reproduced without introducing repulsive components explicitly
in the potential. The relative wave function 
thus shows a nodal behavior in the internal region.

In the present study, we employ a folding-type two-$\alpha$ potential 
that was based on the effective nucleon-nucleon interaction~\cite{Schmid61}
and readjusted in Ref.~\cite{Fukatsu92}.
This potential is expressed in a single Gaussian form
that only includes attractive term.
The calculated energy of $^8$Be is 0.095 MeV,
reproducing the empirical energy.
The explicit form of the potential is a simple Gaussian form:
\begin{align}
  U_{\rm OCM}(r)=-106.1\exp\left(-\frac{r^2}{2.23^2}\right).
\label{2apot2.eq}
\end{align}
This potential is apparently much deeper than
the AB potential of Eq.~(\ref{2apot1.eq}),
and produces the three redundant forbidden states,
which should be removed from all the pairwise wave functions
in the three-$\alpha$ systems.
The present Hamiltonian makes
  the ground- and Hoyle states overbound
  only with the two-$\alpha$ interaction,
  and hence a repulsive phenomenological three-$\alpha$ potential
  is often introduced to adjust these energies to the empirical values.
  Some applications with this potential set
    are given in Refs.~\cite{Kurokawa05,Kurokawa07, Ohtsubo13}.
Because the calculated Hoyle state energy
amounts to no less than 0.78 MeV,
here we newly parametrize a three-$\alpha$ potential
better able to reproduce the empirical Hoyle state energy
0.38 MeV~\cite{Ajzenberg90} for a fair comparison with the AB result.
The explicit form of the potential in MeV is
\begin{align}
U^{(3)}(R)=77.0\exp(-0.12R^2)-10.0\exp(-0.03 R^2).
\end{align}
The calculated Hoyle state energy is 0.34 MeV,
which is close to the empirical energy of the Hoyle state.

\subsection{Correlated Gaussian expansion}
\label{CG.sec}

Let us proceed to construct the wave function of
the $N$-$\alpha$ system, which is expanded by a superposition of
symmetrized correlated Gaussian basis functions~\cite{Varga95, SVM, Suzuki08}
\begin{align}
  \Psi^{(k)}&=\sum_{i=1}^{K} C_i^{(k)}\bar{G}(A_i,\bm{x})\\
  &=\sum_{i=1}^{K} C_i^{(k)}\mathcal{S}\exp\left(-\frac{1}{2}\tilde{\bm{x}}A_i\bm{x}\right),
\end{align}
where $\mathcal{S}$ denotes the symmetrizer
that ensures the symmetry of an identical bosonic system,
$\bm{x}$ is the $(N-1)$-dimensional column vector
composed of a set of the Jacobi coordinate excluding
the center-of-mass coordinate $\bm{x}_N$,
and the tilde denotes the transpose of  the corresponding matrix.
A set of the coefficients $C_i^{(k)}$ can be obtained
by solving the generalized eigenvalue problem
\begin{align}
 \sum_{j=1}^{K} H_{ij} C_{j}^{(k)} = E^{(k)}\sum_{j=1}^K \; B_{ij} C_j^{(k)},
\end{align} 
where
\begin{align}
  H_{ij}&=\left<\bar{G}(A_i,\bm{x})|H|\bar{G}(A_j,\bm{x})\right>
\end{align}
and
\begin{align}
B_{ij}&=\left<\bar{G}(A_i,\bm{x})|\bar{G}(A_j,\bm{x})\right>.
\end{align}
The nonlinear variational parameter $A_i$ is
a positive definite symmetric $(N-1)$-dimensional matrix.
Note that its off-diagonal elements, which control correlations 
among particles, are determined by means of the stochastic variational
method~\cite{Varga95,SVM}.  We follow the setup of Ref.~\cite{LHP20}
to optimize the wave function for the three-$\alpha$ system.
The bound state approximation is applied to the positive
  energy state, which is valid for a state with a narrow decay width~\cite{Horiuchi08,Horiuchi13a,Horiuchi13b}.
The correlated Gaussian approach is flexible enough to describe both
short-range and long-range correlations required in this study.
See~\cite{Mitroy13, Suzuki17} for typical examples that show the power
of this approach.

When we incorporate the deep potential model into the calculations,
we impose the orthogonality condition practically
by using the projection method~\cite{Kukulin78}
that adds the pseudo potential or projection operator,
\begin{align}
  \gamma\sum_{i<j=1}^N\sum_{nlm\in f}\left|\phi_{nlm}(ij)\right>\left<\phi_{nlm}(ij)\right|,
\end{align}
to the Hamiltonian.
One can eliminate the forbidden states variationally
by taking a large $\gamma$ value. 
Here, we adopt the harmonic-oscillator wave functions with
the width parameter $\nu=0.2675$ fm$^{-2}$~\cite{Fukatsu92},
which reproduces the size of the $\alpha$ particle,
as the forbidden states to be eliminated
from the relative motion between the $\alpha$ particles.
We take $\gamma=10^5$ MeV and confirm
that the converged wave functions typically contain
the forbidden state component of relative magnitude
$\approx 10^{-6}$.

\section{Results and discussions}
\label{results.sec}

\begin{figure*}[ht]
\begin{center}
\includegraphics[width=\linewidth]{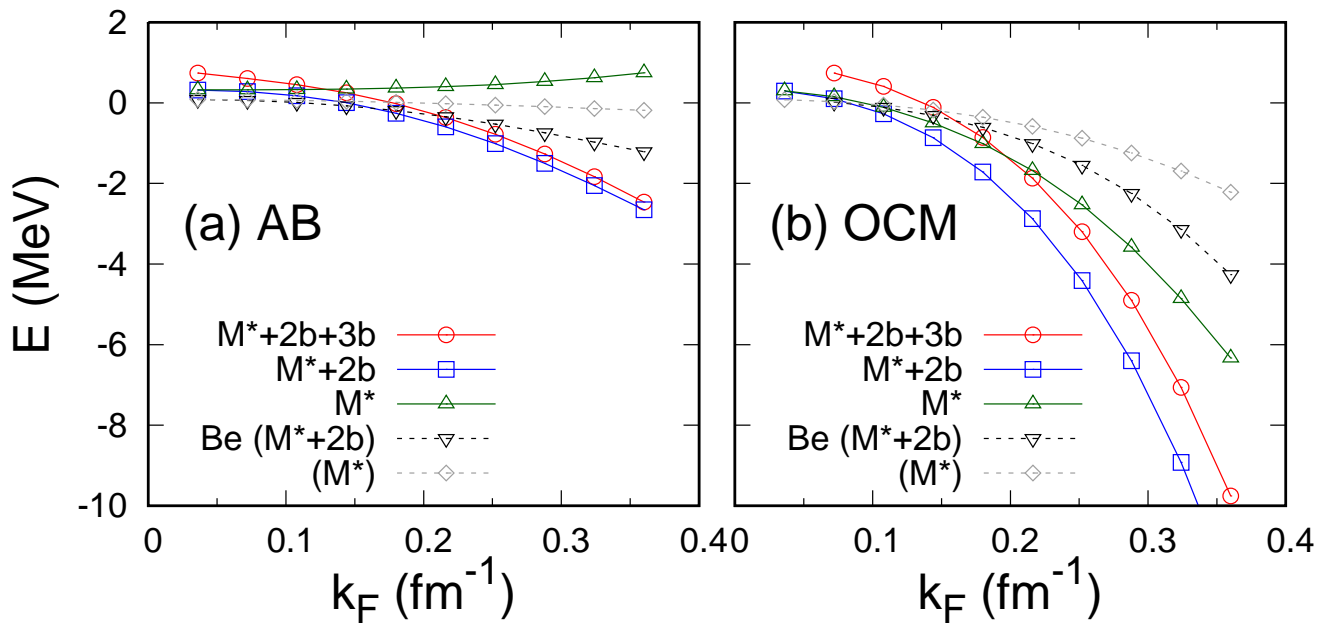}
\end{center}
\caption{Energies of  the three-$\alpha$ system in
  neutron matter calculated as a function of the neutron Fermi momentum
  with different potential models, (a) AB and (b) OCM.
  The calculations including the effective mass alone as medium effects
  are denoted by $M^*$, while those additionally including the induced two-body 
  force and also the induced three-body force are denoted by $M^*+2$b and 
  $M^*+2$b+3b, respectively.
  The results for the two-$\alpha$ system ($^8$Be)
  are also plotted for comparison.
  The lines are guide for the eye.}
\label{fig:energy3}
\end{figure*}

In neutron matter,
the effective mass $M^*$ of an $\alpha$ particle as well as
the induced two- and three-$\alpha$ interactions, changes with
$k_F$~\cite{Nakano20}.
Here we discuss the influence of these medium effects on 
the binding energy of the two- and three-$\alpha$ systems.
More specifically, we analyze the ground state of the two-$\alpha$
system, i.e., $^{8}$Be, and the first excited state 
of the three-$\alpha$ system, i.e., the Hoyle state of $^{12}$C,
both of which exhibit a resonance in vacuum.
We shall show that both the $^{8}$Be and Hoyle states become bound
in the neutron medium of sufficiently large $k_F$.

Figure~\ref{fig:energy3} 
shows the energies of the three-$\alpha$ systems relative to 
the three-$\alpha$ threshold, calculated for AB and OCM 
as a function of the Fermi momentum of the neutron medium $k_F$.
To see the contributions of the induced interactions, we compare 
the energies including the two-body and/or three-body induced interactions
with the one in the absence of the induced interactions.
In general, each energy thus calculated gains as $k_F$ increases
except for the Hoyle state energies only with $M^*$ for AB.
 Note that $M^*$ increases with $k_F$~\cite{Nakano20}, 
 leading to further localization near the potential minima.
 In fact, the results only with $M^*$ contribution clearly
 reflect the properties that the OCM potential only have
 an attractive component
 while the AB potential has repulsive and attractive components
 at short and intermediate distances, respectively.

For the same reason, $\alpha$ particles of larger $M^*$
come closer to each other once the induced two-$\alpha$ 
interaction, which is attractive at short distances 
as shown in Fig.~\ref{fig:induced2bf}, is taken into account.
Then, the induced two-body interaction always plays
a role in gaining the binding energy, which can be seen
in the results allowing for the induced two-body interaction 
($M^*+2$b).
 This is consistent with
a microscopic $\alpha+\alpha+n$ cluster model calculation~\cite{Lyu15}, 
which shows that the $\alpha$-$\alpha$ distance shrinks in $^{9}$Be 
owing to the interaction from the intervening neutron.
On the other hand, the induced three-$\alpha$ interaction
is always repulsive, which leads to increase in
the energy denoted by $M^*+2{\rm b}+3{\rm b}$ as compared with
the one denoted by $M^*+2$b.
The result of OCM ($M^*+2{\rm b}+3{\rm b}$) with $k_F=0.036$ fm$^{-1}$ 
  is not shown
  because no physically stable state is obtained due to
  too strong repulsion of the induced three-$\alpha$ interaction.

While all the above-mentioned tendencies apply to
the two cluster models, quantitative details look very different.
For AB, virtually no contribution from the induced three-body interaction
is found because there is only a negligible wave function amplitude 
in the internal region due to the repulsive component of the AB potential,
which will be shown in the next paragraph.
The ground state of $^8$Be become bound
at $k_F\gtrsim 0.11$ fm$^{-1}$
for AB and $\gtrsim 0.08$ fm$^{-1}$ for OCM.
The Hoyle state becomes bound, i.e.,
the energy is located below the $^{8}$Be energy,
at $k_F\gtrsim 0.22$ fm$^{-1}$ for AB and $\gtrsim 0.16$ fm$^{-1}$ for OCM.
In the OCM case, the condition for binding of the Hoyle state is
 determined by a subtle competition between the
attractive and repulsive contributions
from the induced two- and three-body interactions, respectively.
Incidentally, one can safely ignore
the excited states and dissociation of an $\alpha$ particle
because the excitation energy to the first excited state
in vacuum is far larger than the neutron
Fermi energy at neutron densities of interest here.
Also, we ignore possible increase in the kinetic energy of 
the two- and three-$\alpha$ systems due to the Pauli blocking effect, 
which, in the case of dissolution of $\alpha$ clusters, would
become significant 
when the density of the nuclear medium exceeds 
0.03 fm$^{-3}$~\cite{Roepke20},
again far higher than the medium density considered in this work.
 Although the binding energy of an $\alpha$ particle in such a
 low density medium as considered here
 can be shifted by several MeV~\cite{Roepke09}, 
the formation of the molecular-like states near the threshold
has yet to be affected significantly because the threshold 
energy is also shifted by about the same amount.
We remark in passing that the polaronic quasiparticle energy, 
which is typically of order MeV~\cite{Nakano20}, acts as a shift of the 
binding energy of an $\alpha$ particle, but does not affect the 
formation of weakly bound molecular-like states because the 
threshold energy equally shifts.

\begin{figure*}[th]
\begin{center}
\includegraphics[width=\linewidth]{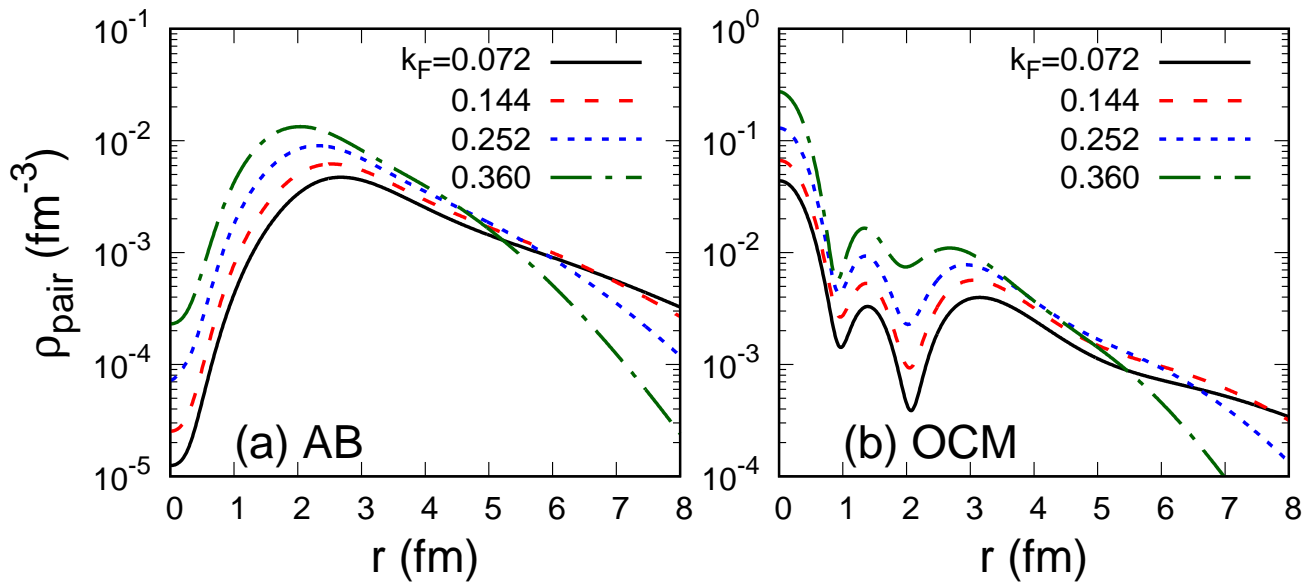}
\end{center}
\caption{Pair density distributions $\rho_{\rm pair}(r)$ of 
  the three-$\alpha$ system in cold neutron matter of various $k_F$ with
  (a) AB and (b) OCM.}
\label{fig:pairdens}
\end{figure*}
 
This model dependence of the system energy comes from
the difference of the internal structure
of the relative wave function between $\alpha$ clusters.
To see such difference explicitly we calculate
the pair density distributions defined by
\begin{align}
\rho_{\rm pair}(r)=\left<\frac{\delta(|\bm{r}_1-\bm{r}_2|-r)}{4\pi r^2}\right>,
\end{align}
where the bracket denotes the expectation value
with the first excited state wave function of the three-$\alpha$ system
and $4\pi \int_0^\infty r^2\rho_{\rm pair}(r)dr=1$.
Figure~\ref{fig:pairdens} compares the results for the pair density distribution
 obtained at various $k_F$.   For AB,
the amplitude of the wave function is strongly suppressed
due to the repulsive potential component at short distances,
$\lesssim 2$ fm,  while the peak of the amplitude, located near the
potential minimum that arises from $U^{(3)}$, naturally increases with 
$k_F$ or $M^*$.
In the OCM results, on the other hand, an
oscillatory behavior is found at distances $\lesssim 3$ fm
due to the orthogonality condition to the Pauli forbidden states.
Since a significant amount of amplitude is present in such an 
internal  region, the wave function in this region
is strongly modified as the Hamiltonian changes.
For larger $k_F$ or $M^*$, the amplitude
of the internal wave function becomes larger,
which is natural considering that
heavier $\alpha$ particles are more difficult to move
near the OCM potential minimum of zero separation.

\begin{figure*}[ht]
\begin{center}
\includegraphics[width=\linewidth]{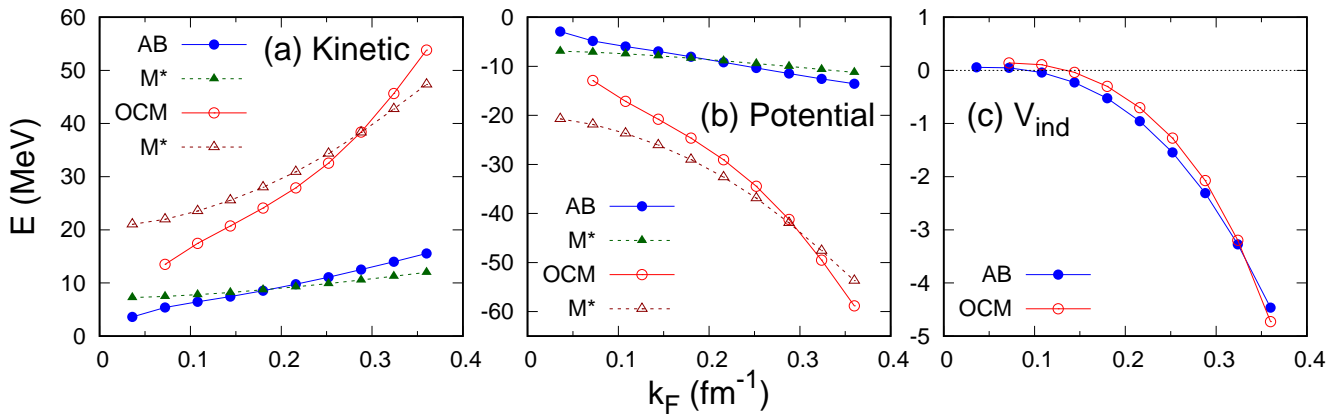}
\end{center}
\caption{Decomposition of the total three-$\alpha$ energy into
  the kinetic term (a), the direct potential term (b), 
  and the induced interaction term (c), which are calculated
  with AB and OCM as a function of the neutron Fermi momentum. 
  See text for details. The thin horizontal line in the panel (c)
  indicates zero.}
\label{fig:components}
\end{figure*}

We conclude this section by examining how the difference in the
pair density distribution between AB and OCM is reflected in
the expectation values of the Hamiltonian terms.
Figure~\ref{fig:components} displays
decomposition of the total energy into the
contributions of the kinetic, direct interaction, 
and induced interaction terms.
Since the OCM wave function has its internal amplitude
disturbed drastically by the medium,
the expectation value of the kinetic energy
rapidly increases as the $k_F$ increases for OCM, as can be
seen from Fig. ~\ref{fig:components} (a).
This energy cost is dominated by
the energy gain from the direct term
$\left<\sum_{ij}U_{ij}^{(2)}+U^{(3)}\right>$
as plotted in Fig.~\ref{fig:components} (b), which is in turn
controlled by the two-body OCM potential responsible for
the zero-separation potential minimum.
For the AB model, the same kind of behavior of 
both terms occurs, but the medium effects are suppressed 
due to the repulsive nature of the AB potential 
at short distances.  Finally,
Fig.~\ref{fig:components} (c) compares
the sum of the expectation values
from the induced two- and three-body interactions 
$\left<\sum_{ij}V_{{\rm eff};ij}^{(2)}+V_{\rm eff}^{(3)}\right>$ 
(denoted by ${\rm V_{ind}}$) between AB and OCM.
In either case, the contribution of the induced three-body force is 
about one or two orders of magnitude smaller than that of 
the induced two-body force.  The model dependence of the induced 
interaction term is appreciable at large $k_F$, a feature that stems 
from the difference in the amplitude of the wave function 
near zero separation via the induced two-body force.
At small $k_F$, the expectation values of the induced
  two and three-$\alpha$ interactions become positive,
  where the magnitude of the repulsive induced three-$\alpha$ interaction
  is larger than that of the induced two-$\alpha$ interaction.
  This confirms why we do not find any stable Hoyle state
  for the OCM result with $k_F=0.036$ fm$^{-1}$.

The decomposition in the absence of the induced two- and three-body 
interactions is also plotted in Figs.~\ref{fig:components} 
(a) and (b) as denoted by $M^*$.
We see that both the kinetic and direct interaction terms
almost follow the full calculations.
Since the contributions from the induced interactions
are minor, i.e., one order of magnitude smaller than the expectation
values of the kinetic and direct interaction terms,
the $k_F$ dependence is predominantly determined by the Hamiltonian
 in the absence of the medium effects except the effective mass
correction.
The modeling of the  $\alpha$ cluster structure is more essential
than the medium-induced interactions to describe the $k_F$ 
dependence of the properties of the three-$\alpha$ system
in cold neutron matter.

In the present study, 
the two- and three-$\alpha$ systems, once being bound,
have an infinitely long lifetime.  This is because
we have assumed that each $\alpha$ particle is robust 
in dilute, cold neutron medium 
and that possible medium effects come into our calculations only 
through the effective mass and in-medium interactions. 
In order to evaluate the lifetime of the two- and three-$\alpha$ bound states
in the present cluster picture, we have to take into account a multiple scattering process 
among $\alpha$ particles and surrounding neutrons, e.g., in few-body 
T-matrix approach.  Such a process has not been considered in the present study.  
We expect, however, that the resultant width (the inverse lifetime) 
of these bound states would be negligibly small compared to the in-medium energy 
shift of each $\alpha$ particle, because, as was found in Ref.\ \cite{Nakano20}, 
the decay process from a single polaronic $\alpha$ particle to a bare $\alpha$
particle and neutrons is kinematically suppressed due to the neutron 
Fermi degeneracy at low temperature.  We expect that a similar mechanism 
works also for the two- and three-$\alpha$ particle systems; 
the broadening is not so large as to lose the cluster picture.

\section{Conclusion and future prospective}
\label{conclusion.sec}

The possibility that normally resonant two- and three-$\alpha$ systems
 become bound in cold neutron matter has been pointed out
for the first time by combining precise quantum-mechanical calculations 
with a polaron picture of $\alpha$ particles.
We have examined two standard $\alpha$-cluster models that take into account
the Pauli principle in a different way,  i.e., via the Pauli potential
and the orthogonality condition to the Pauli forbidden bound states.
We have shown that the ground state of $^8$Be
and the Hoyle state can be bound at $k_F\gtrsim 0.08$--0.11 fm$^{-1}$ and
$k_F\gtrsim 0.16$--0.22 fm$^{-1}$, respectively, for the two models.
The presence of these light nuclear ingredients as bound states
would give a significant impact on the modeling of matter in
stellar collapse and neutron star mergers and also affect reaction 
rates for nucleosynthesis therein.

It is interesting to note that
the in-medium attraction discussed in this work
has to be realized in finite nuclear systems,
e.g., Be and C isotopes, where the $\alpha$ cluster
structure is well developed. See, e.g., Ref.~\cite{Enyo15} and
references therein.
Isotope dependence of the structure
of $2\alpha+Xn$ and $3\alpha+Xn$ systems would drop a hint
 at the stability of $\alpha$ clusters in cold neutron matter.
As this is just the first evaluation, for simplicity,
we ignore the distortion of an $\alpha$ particle and
the Pauli constraint of the relative wave function of
$\alpha$ particles by the surrounding neutron matter.
The latter contribution would work as repulsion and 
might counteract the stability of the 
`bound' $^8$Be and Hoyle states.
It would be desired to develop a model that includes
such explicit correlations from the neutron medium
by starting from the nucleon 
degrees of freedom.

Moreover, finite temperature effects would be important in 
core-collapse supernovae and neutron star mergers.
Although we use the zero-temperature results for in-medium excitation 
properties of a single $\alpha$  particle and induced two- and 
three-$\alpha$ interactions, the description of such in-medium properties 
can be extended to the finite-temperature case along the theoretical 
developments in cold atom physics~\cite{Hu18,Tajima18,Liu19}. 
Works in these directions are underway and will be reported
elsewhere.

\acknowledgments

This work was in part supported by JSPS KAKENHI
Nos.\ 17K05445, 18K03635, 18H01211, 18H04569, 18H05406, 
and 19H05140, and the Collaborative Research Program 2021,
Information Initiative Center, Hokkaido University.

\end{document}